\documentstyle[twoside,fleqn,espcrc2,epsfig]{article}

\def\msbar{${\rm{\overline{MS}}}$} 

\newcommand{\AmS}{{\protect\the\textfont2
  A\kern-.1667em\lower.5ex\hbox{M}\kern-.125emS}}

\newcommand{\ee}{\end{equation}}
\newcommand{\bea}{\begin{eqnarray}}
\newcommand{\eea}{\end{eqnarray}}

\title{Charm Production and Fragmentation in Charged Current DIS}

\author{S.\ Kretzer\address{Institut f\"ur Physik, Universit\"at Dortmund \\ 
       D--44221 Dortmund, Germany 
       (e-mail: kretzer@hal1.physik.uni-dortmund.de) }%
        \thanks{Participation at the {\it{QCD 98}} financed by the  
{\it{Graduiertenkolleg 'Erzeugung und Zerf\"alle von Elementarteilchen'}}
of the {\it{Deutsche Forschungsgemeinschaft}} at the {\it{Universit\"at Dortmund}}.
Work supported in part by the {\it{Bundesministerium f\"{u}r Bildung, 
Wissenschaft, Forschung und
Technologie}}, Bonn. }}
       
\begin{document}

\begin{abstract}
In charged current deep inelastic scattering charm is dominantly produced
in scattering events on strange quarks, thereby allowing for an experimental
determination of the nucleon's strange sea density. A measurement of the energy 
spectrum of final state charm fragments (D-mesons) determines 
the charm fragmentation function at spacelike scales considerably below typical
$e^+e^-$ c.m.s.\ energies. NLO corrections to the naive $s\rightarrow c$
parton model production picture are important and well understood.  
\end{abstract}

\maketitle

\section{Parton Model Expectations}

Charm production in CC DIS is dominated by an $s \rightarrow c$ transition at 
the virtual $W$-boson vertex. The $d \rightarrow c$ background 
is sizable at large $x$ where the Cabibbo suppression is balanced 
by the valence enhancement of $d_v$. Since $d_v$ is well known I will not 
consider $d \rightarrow c$ and assume a vanishing Cabibbo angle for simplicity.
The results presented here are, however, not affected by this choice.
The LO production cross section of charmed hadrons
\begin{equation}
\label{simple}
\frac{d \sigma_{LO}}{dx\ dy\ dz} \propto s(\chi) D_c(z) 
\end{equation}
manifests the obvious possibility to extract the nucleon's strange sea
density $s(\chi)$ and the charm fragmentation function $D_c(z)$
from experimental data \cite{exp}. 
In Eq.\ (\ref{simple}) $x$ and $y$ are the  
standard inclusive DIS observables and $z\equiv p_{H_c}\cdot p_N/q\cdot p_N$;
$p_{H_c}$, $p_N$ and $q$ being the momentum of the charmed hadron, the 
target nucleon and the mediated gauge boson, respectively.
In the target rest frame $z$ reduces to the charmed hadron energy 
$E_{H_c}$ scaled to its maximal value $\nu=q_0$. In the massless parton model
the fractional momentum
$\chi$ of the struck strange quark reduces to the Bjorken variable $x$. 
For theoretical predictions
to be reliable one has of course to consider the NLO of the QCD perturbation
series where $W^\ast s \rightarrow c g$ (incl.\ virtual corrections)
and $W^\ast g \rightarrow c {\bar{s}}$ contributions enter the game.
The NLO terms mix quark and gluon initiated contributions and induce
a scheme dependence due to the necessity of handling collinear divergencies
arising from on-shell strange quark propagators. Most prominent choices are
the dimensionally regularized ($m_s=0$)
\msbar\ scheme and the massive ($m_s \neq 0$) ACOT
formalism \cite{acot} for heavy quarks. 

In principle also $\Delta s$ 
seems measurable at a polarized HERA setup \cite{masch} and the 
corresponding NLO framework is underway \cite{krstr}. 
 

\section{NLO framework}

At NLO the production cross section is no longer of the simple factorized
form of Eq.\ (\ref{simple}) and double convolutions 
(symbol $\otimes$ below)  have to be considered \cite{gkr}.
However, to a reasonable approximation
\begin{eqnarray} \nonumber
d \sigma_{NLO} &=& \left( \left[ s \otimes d{\hat{\sigma}}_s 
+ g \otimes d{\hat{\sigma}}_g \right ] \otimes D_c \right) 
(x,z,Q^2) \\ \nonumber
&\equiv& d \sigma_{LO}\ K(x,z,Q^2) \\ \nonumber
&\propto& s(\chi)\ D_c(z)\ K(x,z,Q^2) \\
&\simeq& s(\chi)\ {\cal{D}}_{x,Q^2}[D_c](z) 
\label{kfactor}
\end{eqnarray}
holds
\begin{figure}[t]
\vspace*{-0.5cm}
\hspace*{-1.25cm}
\epsfig{figure=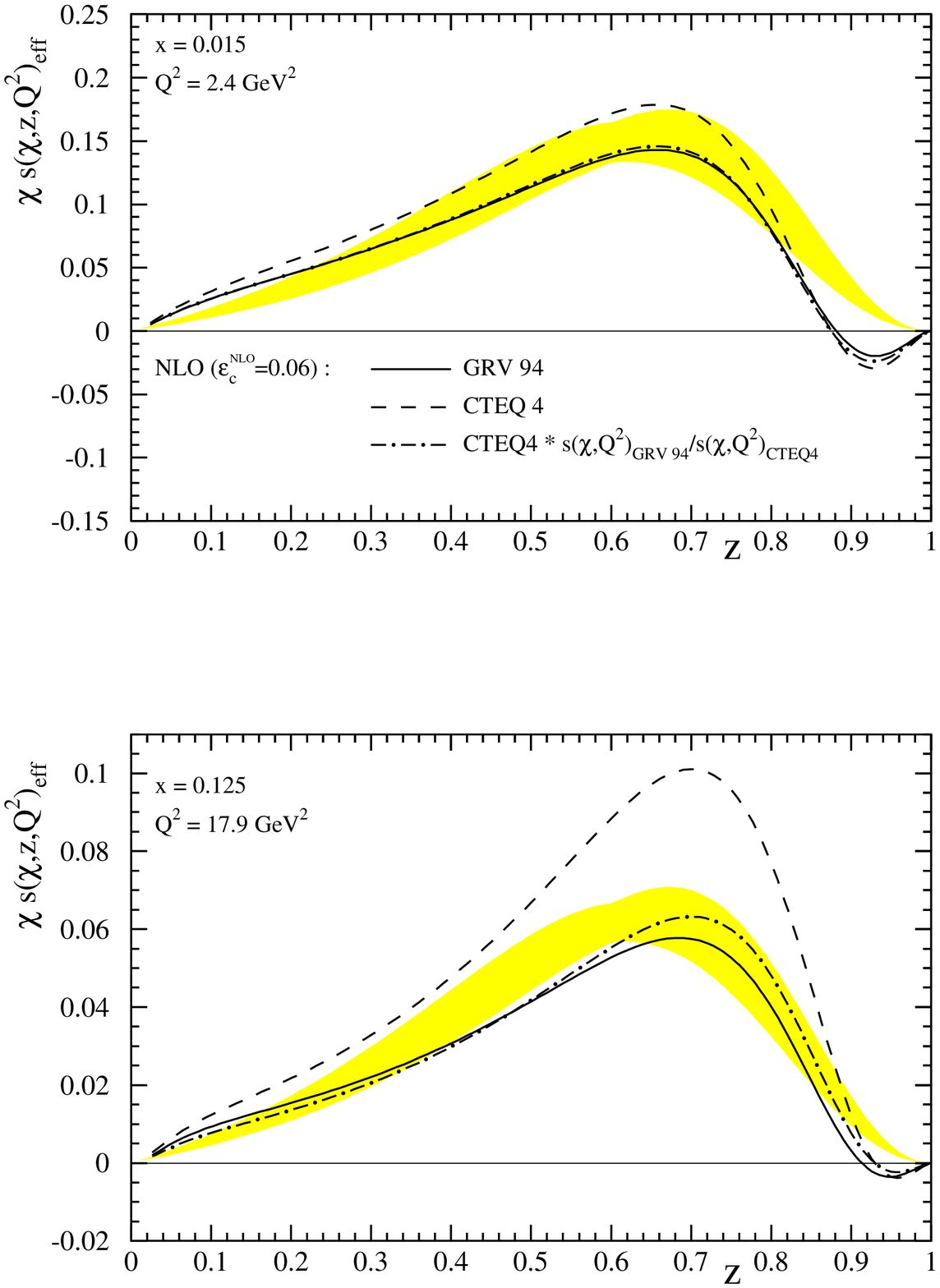,width=9cm}
\vspace*{-2.5cm}
\caption{$s_{eff}$ equals the charm production cross section in 
Eq.\ (\ref{kfactor}) up to a constant of normalization. The shaded band
represents a parametrization of CCFR data and the curves show theoretical
predictions using GRV94 (solid) and CTEQ4 (dashed). 
For the dot-dashed
curves the normalization of the CTEQ4 prediction has been changed by
the factor given in the legend.}
\end{figure}
also at NLO accuracy within experimental errors
and for the limited kinematical range of present data
on neutrinoproduction of charm. In Eq.\ (\ref{kfactor}) the approximate 
multiplicative factor ${\cal{D}}$ absorbs the precise $K$-factor $K(x,z,Q^2)$
obtained from a full NLO QCD calculation \cite{gkr}. ${\cal{D}}$
is {\it{not}} a simple universal fragmentation
function but a nontrivial process-dependent functional 
which is, however, mainly sensitive on $D_c$ and shows little sensitivity on
the exact parton distributions considered. 
The occurrence of $x,Q^2$ and $z$ in Eq.\ (\ref{kfactor}) as indices and 
as a functional argument, respectively, reflects the fact that the dependence
on $x$ and $Q^2$ is much weaker than is on $z$. 
Eq.\ (\ref{kfactor}) tells us that $s(\chi)$ fixes the normalization of
$d \sigma$ once $K$ is known. On the other hand $K$ (or ${\cal{D}}$) can be
computed \cite{gkr} from $D_c$ with little sensitivity 
on $s(\chi)$, 
such that $s(\chi)$ and $D_c(z)$ decouple in the production
dynamics and can be simultaneously extracted. 
This point can be clearly inferred from Fig.\ 1 where it is shown
for CCFR \cite{exp} (fixed target) kinematics
that the NLO calculations \cite{gkr} using 
GRV94 \cite{grv94} (solid) and CTEQ4 \cite{cteq4} (dashed)
strange seas can be brought into good agreement by a mere
change of the normalization given by the ratio 
$s_{GRV}(\chi)/s_{CTEQ4}(\chi)$ (dot-dashed).
The freedom in realizing Eq.\ (\ref{kfactor}) in distinct NLO schemes, 
e.g.\ \msbar\ or ACOT, might a priori lead to an ambiguity in measuring the
nucleon's strangeness. From Fig.\ 2 one can, however, see that the scheme
dependence is small \cite{kschie1}
for the inclusive (z-integrated) charm structure
function $F_2^c$. This result does also hold for the semi-inclusive 
structure functions entering Eq.\ (\ref{kfactor}) \cite{kschie2},
as can be judged comparing the $m_s=0$ and $m_s=500 {\rm{MeV}}$ curves in
the upper half of Fig.\ 4 below. 
\begin{figure}[t]
\vspace*{-0.5cm}
\hspace*{-1.25cm}
\epsfig{figure=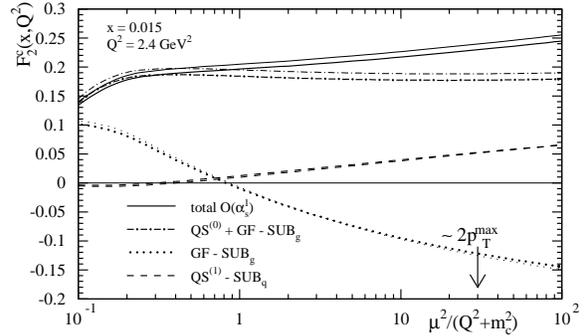,width=9cm}
\vspace*{-2.5cm}
  \caption{The inclusive structure function $F_2^c$ for charm production
in CC DIS (solid lines) within \msbar\ (thick lines) and ACOT (thin lines)
using $m_s=500 {\rm{MeV}}$. Also shown are the individual quark scattering
and gluon fusion components which contribute to the structure function
and which are regularized by subtraction terms. $\mu$ is the arbitrary factorization
scale.}
\end{figure}

\section{Strange Sea}

The wide spread of available strange sea densities is illustrated 
by some of its representatives \cite{grv94,cteq4,exp} in Fig.\ 3.
It is especially interesting to note the difference -not only in
size but also in shape- of the characteristically steep strange sea
of GRV94 which builds up entirely through a renormalization group
resummation of $g \rightarrow s {\bar{s}}$ splittings from a low 
resolution scale $\mu^2 \sim 0.3 {\rm{GeV}}^2$ and the conventional
strange seas of CTEQ4 and CCFR which comprise an additional nonperturbative
input component at larger $x\sim 0.1$. 
Comparing the solid (GRV94) and the dashed (CTEQ4) curves in Fig.\ 1 with the 
parametrization \cite{gkr} of CCFR production data a purely radiative strange 
sea seems to be favored over larger nonperturbative inputs which overshoot data 
at larger $x$. This conclusion is awaiting further experimental
analyses and it would be helpful to have published production data at hand
for future investigations.
\begin{figure}[t]
\vspace*{-0.5cm}
\hspace*{-1.25cm}
\epsfig{figure=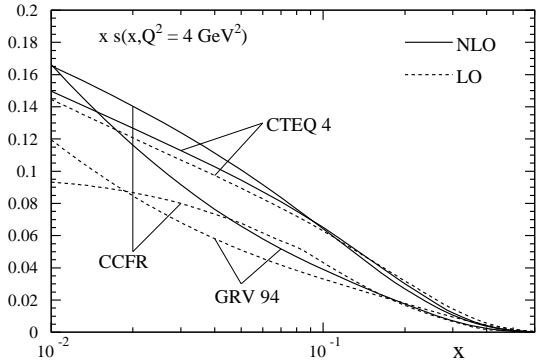,width=9cm}
\vspace*{-2.cm}
  \caption{Several leading and next-to-leading order strange sea densities.}
\end{figure}

\section{Charm Fragmentation Function (FF)}

Since the cross section in Eq.\ (\ref{simple}) is directly proportional
to $D_c$ at LO accuracy neutrinoproduction of charm can give valuable
information on the charm FF complementary
to that from one charmed hadron inclusive $e^+e^-$ annihilation.
Especially a test of the universality of the charm fragmentation function
is an important issue \cite{klre}. In Fig.\ 1 a scale-independent
Peterson \cite{peterson} FF with a hardness parameter of $\varepsilon_c=0.06$
has been used. This choice seems to be compatible with the neutrino data
band. On the other hand a distinctly harder value of $\varepsilon_c=0.02$
has been obtained in \cite{cagre} from LEP and ARGUS $e^+e^-$ data, see also 
\cite{bkk} for related analyses. 
If the fit of \cite{cagre} is evolved down to fixed target energies
it is incompatible with the parametrized neutrino data in Fig.\ 1.
\begin{figure}[t]
\vspace*{-0.1cm}
\hspace*{-1.25cm}
\epsfig{figure=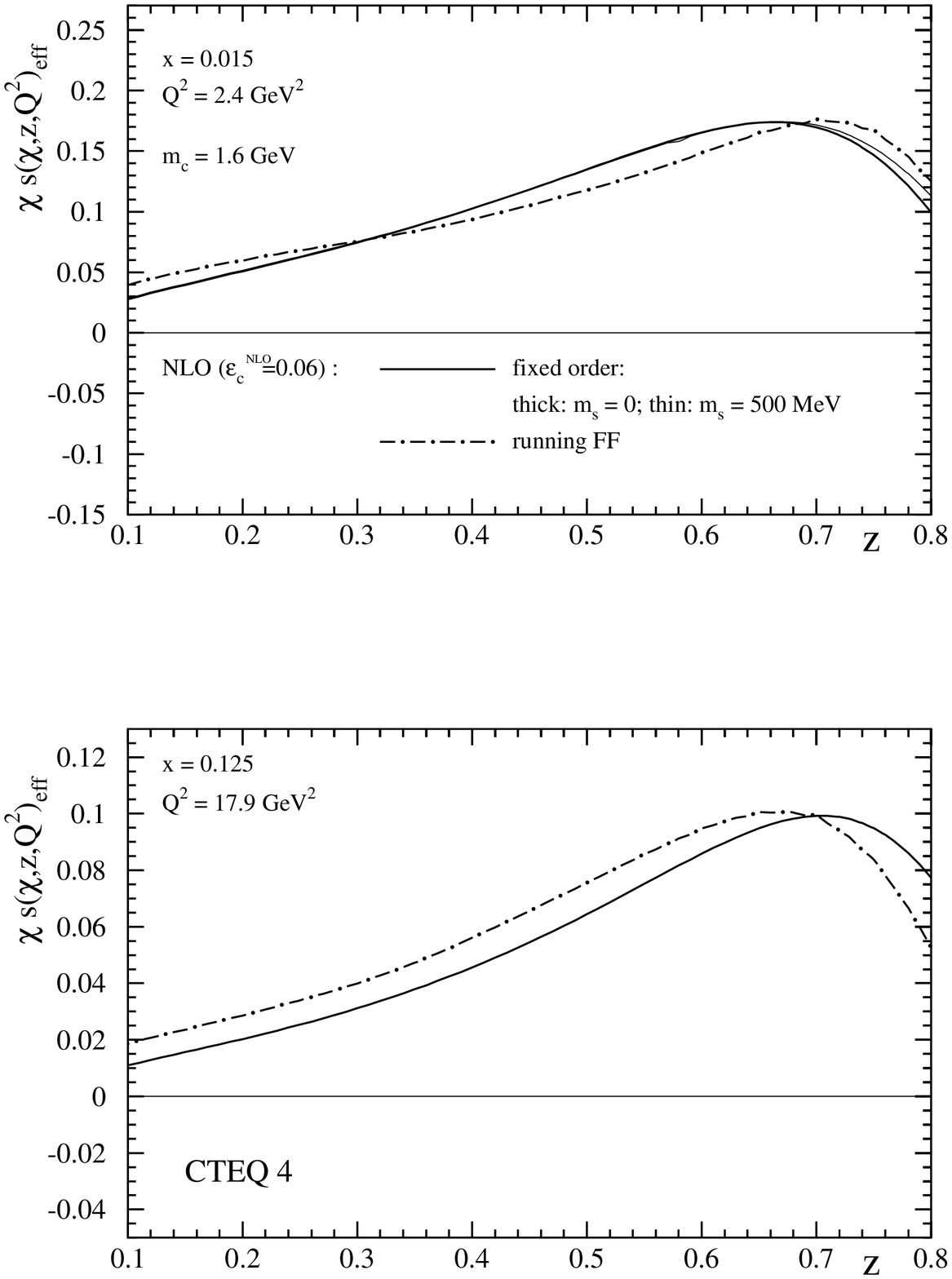,width=9cm}
\vspace*{-2.5cm}
  \caption{Scheme dependence of the production cross section in 
Eq.\ (\ref{kfactor}). For the dot-dashed curves the quasi collinear
logs $\ln m_c^2$ of the fixed order results (solid) have been resummed into
a running of the fragmentation function. In the upper half possible finite 
$m_s$ effects are furthermore considered.}
\end{figure}
A point which might influence the extraction of a universal
FF from neutrinoproduction is the scheme dependence in handling
final state quasi-collinear logarithms $\ln (Q^2/m_c^2)$. 
Along the lines of \cite{melnas}
they may be resummed into a running of the charm FF as has been done 
in \cite{cagre}
or they may be kept at fixed order as in \cite{gkr} using a scale-independent
FF.
In Fig.\ 4 we examine such resummation effects for 
CCFR kinematics. We use the same Peterson FF with 
$\varepsilon_c=0.06$ once for a fixed order calculation
\cite{gkr} (solid lines) and once as the
nonperturbative input part of the running 
$c\rightarrow D$ FF (dashed curves).
We note that towards intermediate scales around 
$Q^2\sim 20 {\rm{GeV}}^2$
one begins to see the softening effects of the resummation
which are enhanced as compared to the fixed order calculation. However,
as one  would expect at these scales, the resummation effects are 
moderate  and cannot explain the descrepancy between the $\varepsilon_c$
values. However, charm fragmentation at LEP
has been measured by tagging on $D^\ast$'s, whereas neutrinoproduction
experiments observe mainly $D$'s through their semileptonic decay-channel
(dimuon events). ARGUS \cite{argus} and CLEO \cite{cleo} data at 
$\sqrt{s}\simeq 10 {\rm{GeV}}$ indeed show \cite{pdg} 
a harder energy distribution of $D^\ast$'s compared
to $D$'s. It seems therefore to be possible within experimental accuracy to observe 
a nondegeneracy of the charm fragmentation functions into the lowest
charmed pseudoscalar and vector mesons.  
We note that an $\varepsilon_c$ value around $0.06$ which is in agreement
with neutrino data on $D$-production
is also compatible with the $D$ energy spectrum measured
at ARGUS where the evolution may be performed either using fixed order
expressions in \cite{nasweb} or via a RG transformation along the lines of
\cite{melnas,cagre}. If forthcoming experimental analyses should confirm our 
findings the lower decade $m_c(\sim 1{\rm{GeV}})\rightarrow$ ARGUS$(10{\rm{GeV}})$
may be added to 
the evolution path ARGUS$(10{\rm{GeV}})\rightarrow$LEP$(M_Z)$ paved in 
\cite{cagre} for the charm FF.

\subsection*{Acknowledgements}
I am grateful to E.\ Reya and M.\ Gl\"uck for advice and  useful discussions.
I want to thank I.\ Schienbein for a fruitful collaboration on many of
the results presented here.

\end{document}